\documentclass[a4paper,10pt]{article}
\usepackage{amsmath}
\usepackage{amsfonts}
\usepackage{amssymb}
\usepackage{graphicx}
\begin{document}
\author {S.K.Tripathy $^1$ and K.L.Mahanta$^2$ \\
1  Department of Physics,\\
Indira Gandhi Institute of Technology, Sarang,\\
Dhenkanal, Odisha-759146, INDIA.\\
e-mail: tripathy\_ sunil@rediffmail.com\\
2 Department of Mathematics,\\
C. V. Raman College of Engineering \\
Mahura, Janla, Bhubaneswar, Odisha, INDIA.\\
e-mail: kamal2\_m@yahoo.com}

\title  {Cosmic Acceleration and Anisotropic models with Magnetic field}

\maketitle

\begin{abstract}
Plane symmetric cosmological models are investigated with or without any dark energy components in the field equations. Keeping an eye on the recent observational constraints concerning the accelerating phase of expansion of the universe, the role of magnetic field is assessed. In the absence of dark energy components, magnetic field can favour an accelerating model even if we take a linear relationship between the directional Hubble parameters. In presence of dark energy components in the form of a time varying cosmological constant, the influence of magnetic field is found to be limited.
\end{abstract}


\section{Introduction}
Observations from distant type Ia Supernovae confirm that presently the universe is undergoing an accelerated phase of expansion \cite{Reiss98,Reiss04, Perl98,Knop03, Reiss07}. These data led to the development of a lot of novel ideas and solutions.  The accelerated expansion is believed to be due to an exotic form of energy, known as dark energy. Observations of large scale structure and cosmic microwave background (CMB) also provide strong evidence in favour of dark energy \cite{Sperg07, Blan06}. The presence of dark energy with a negative pressure is confirmed with additional evidences from observations of X-ray clusters \cite{Allen04} and Baryon Acoustic Oscillations (BAO) \cite{Ein05}. In recent works by Sullivan et al \cite{Sul11} and Suzuki et al. \cite{Suz12} cosmic acceleration with dark energy components has gained much support and a tighter constraint has been put on the dark energy equation of state.  A lot of  recent observational data suggest that dark energy is the dominant component in the mass-energy budget of the universe with a share of $68.3 \%$. Dark matter and  baryonic matter comprise only $26.8\%$ and $4.9 \%$ respectively \cite{Ade14, Ade13a,Ade13b,Ade13c}. Dark energy provides a strong negative pressure giving rise to an anti gravity effect that drives the acceleration (for recent reviews see \cite{Copeland06,Caldwell09,Silvestri09, Astier12} and references therein). In the framework of general relativity a fluid with a static or almost static density may cause such a cosmic acceleration. Dark energy refers to such a hypothetical fluid \cite {Astier12}. The simplest and natural candidate for  dark energy is a cosmological constant in classical FRW model or the $\Lambda$ dominated Cold Dark Matter ($\Lambda$CDM ) model. But the cosmological constant faces  many  serious problems like the fine tuning problem, coincidence problem etc.( see \cite{Carol01,Wein89} for reviews on cosmological constant problem). Many other dark energy models have been proposed in recent times with alternative candidates such as variable cosmological constant, a canonical scalar field like quintessence models \cite{Ratra88}, a phantom field, a scalar field with negative kinetic term \cite{Caldwell02}, ghost condensate \cite{Piazza04} or k-essence \cite{Chiba00}. On the other hand, dynamical dark energy is effectively described by a modification of geometrical part of Einstein-Hilbert action or modification of gravity  by using functions of curvature scalar ($f(R)$ gravity models \cite{Nojiri03, Felice10}), of Gauss-Bonnet invariant \cite{Nojiri05} or higher derivatives of the action \cite{Nojiri05a}, holographic properties \cite{Li04} etc. However, it remains a challenging task to understand the accelerated cosmic expansion and thus the nature  and origin of dark energy is still elusive.

The standard cosmological model ($\Lambda$CDM)based upon the spatial isotropy and flatness of the universe is consistent with the data from  precise measurements of the CMB temperature anisotropy \cite{Hinshaw09} from Wilkinson Microwave Anisotropy Probe(WMAP). However, the $\Lambda$CDM model suffers from some anomalous features at large scale such as (i)observed large scale velocity flows than prediction , (ii) a statistically significant alignment and planarity of the CMB quadrupole and octupole modes and (iii) the observed large scale alignment in the quasar polarization vectors \cite{Antoniou12}. Recently released Planck data \cite{Ade14, Ade13a, Ade13b,Ade13c} show a slight red-shift of the primordial power spectrum from the exact scale invariance. The anisotropic parameter $g_*$ in the power spectrum of curvature perturbation $\varsigma $ with broken statistical isotropy, $P_{\varsigma}(\vec{k} )=P_{\varsigma}^{0}(\vec k)\left(1+g_{*}Cos^{2}\theta_{\vec{k},v}\right)$ \cite{Acker07,Ohashi13},   describing the deviation from the isotropic behaviour is constrained to be $g_*=0.29\pm 0.031$ from WMAP data \cite{Groen10} and $|g_*|\ge 0.5$ from Planck data \cite{Ohashi13}. It is clear from the Planck data that, $\Lambda$CDM model does not fit well to the temperature power spectrum at low multipoles \cite{Ade13a}. Also, precise measurements from WMAP predict aymmetric expansion with one direction expanding differently from the other two transverse direction at equatorial plane \cite{Buiny06} which signals a non trivial topology of the large scale geometry of the universe \cite{Wata09}. In order to address the issue of the smallness in the angular power spectrum of the temperature anisotropy, plane symmetric models have been proposed in recent times \cite{Campa06, Campa07, Campa09, Gruppo07}. Plane Symmetric or Locally Rotationally Symmetric Bianchi type I (LRSBI) models have  also been studied in different context to address different issues of cosmology \cite{Tripathy08,Tripathy09, Tripathy09a, Tripathy10,Tripathy13a, Tripathy13b, Rod08, Sharif11, Mohanty08}. These models are  more interesting in the sense that they are more general then FRW models.

The universe contains highly ionized matter and therefore, magnetic field plays an important role in the description of its dynamics and energy distribution. It plays a crucial role in star formation, solar and stellar activity, pulsars, accretion  disks, formation and stability of jets and stability of galactic disks \cite{Pino06}. Strong magnetic field may be created due to adiabatic compression in cluster of galaxies. Cosmic anisotropies may also be attributed to the large scale magnetic fields. In anisotropic models they could alter the particle creation rates and can affect the rate of expansion. The problem of the origin and possible amplification of cosmic magnetic field has been discussed by many authors \cite{Grasso01, Giovannini04, Giovannini08}. The origin of cosmic magnetic fields can be attributed to primordial quantum fluctuations. The large scale galactic, intergalactic and super cluster magnetic fields are of the order of $10^{-6}$Gauss to $10^{-11}$Gauss with correlation from $100 Kpc$ to several Mpc to the extent that they are originated from scalar and possibly gauge field fluctuations after exciting the inflation. Their seeds may be in the range $10^{-18}-10^{-27}$Gauss or less \cite{Grasso01, Giovannini04, Giovannini08, Andrianov08}. Magnetic field with an amplitude of $10^{-8}-10^{-9}Gauss$ is believed to leave traces on CMB.

During the phase transition in the early universe with spontaneous symmetry breaking, strings arise as a random network of line-like defects. One dimensional strings are believed to be some topological stable defects, like magnetic monopole and domain walls. Massive closed loops of string serve as seeds for the formation of large scale structures like galaxies and cluster of galaxies \cite{Vilen81}.  While matter is accreted onto loops, they oscillate violently and lose their energy by gravitational radiation. Therefore, they shrink and disappear \cite{Vilen87}. This radiation coming as a signal of early epoch cosmic strings may be detectable in experiments for gravitational waves. Earth bound Laser Interferometer Gravitational-Wave Observatory (LIGO) and space based Laser Interferometer Space Antenna (LISA) to be engaged in detection of gravitational waves may detect the signals from cosmic strings. Cosmic strings are believed to induce temperature anisotropy in CMB. The mean angular power spectrum of string-induced CMB temperature anisotropies can be described by a power law which suggests that a nonvanishing string contribution to the overall CMB anisotropies may become the dominant source of fluctuations at small angular scales \cite{Fraisse08}.

In the context of the present observational data concerning the cosmic acceleration and anisotropy in the temperature power spectrum, it is interesting to investigate the role of magnetic field in anisotropic models in getting an accelerated phase. Motivated by this idea, in the present work, we have investigated LRSBI models in the frame work of Barber's Self Creation Cosmology (BSCC)\cite{Barber82}. Barber \cite{Barber82} proposed two continuous self creation theories modifying Brans-Dicke \cite{Brans61} theory and general theory of relativity (GR). In BSCC, the conservation of energy-momentum is relaxed and the matter universe is created out of self contained gravitational, scalar and matter fields \cite{Barber06}. Brans \cite{Brans87} pointed out that the first theory of Barber violates the equivalence principle and also is in disagreement with experiment and hence the first theory was rejected with gross internal inconsistency.  However, the second theory is an interesting cosmological model and passes all experimental tests to date \cite{Barber10}. The BSCC has been shown by Barber \cite{Barber06} to yield a concordant solution that does not require inflation, exotic non-baryonic dark matter or dark energy to fit observational constraints. There exists a conformal equivalence between this theory and canonical general relativity and hence the two theories predict identical experimental results in standard tests. In his paper \cite{Barber06}, Barber has shown that BSCC has passed all the classical tests such as the deflection of light by sun, the gravitational red shift of light and the precession of perihelia of the orbit of Mercury, the time delay of radar echoes passing the sun, the precession of a gyroscope in earth orbit, the binary pulsar PSR 1913+16. Also he has suggested some definitive tests  that distinguish between BSCC and GR. Barber's Self Creation Cosmology has generated a lot of interest in recent times and has been widely studied \cite{Pradhan02,Singh12, Katore11, Belinchon13, Ramirez13}.

We consider the universe to be filled with an anisotropic pressure less fluid with a cloud of cosmic strings embedded in a magnetic field. Since the experimental evidences for cosmic strings are poor \cite{Ade13d}, the magnetic field can be chosen to be aligned along any arbitrary direction. In the present work, we consider the magnetic field to be aligned along the direction of the cosmic strings. A linear relationship between the directional Hubble parameters is assumed. The pressure anisotropy in the cosmic fluid comes from the presence of cosmic strings and magnetic field. In our earlier works \cite{Tripathy08,Tripathy09, Tripathy09a, Tripathy10}, we have investigated plane symmetric anisotropic models with cosmic strings, bulk viscosity or magnetic strings in GR. However, in the present work, we extend those works to investigate the role of magnetic field towards cosmic acceleration in the framework of Barber's Self Creation Cosmology. In order to assess the role of magnetic field in getting an accelerating model, the deceleration parameter in the presence and in the absence of magnetic field is calculated. 

\paragraph{}The organisation of the paper is as follows. In Sect-2, dynamics of the LRSBI model is discussed for a cloud of cosmic strings in presence of magnetic field. In order to get a clear picture about the role of magnetic field on the properties of the model, we have also presented the results in the absence of magnetic field. In Sect-3, we have incorporated dark energy components in the form of a time varying cosmological constant in the field equations and investigated the role of magnetic field and dark energy component in getting an accelerated phase. At the end, conclusion of the present work is presented in Sect-4.

\section{Anisotropic Cosmological model in presence of Magnetic field}
We consider the plane symmetric LRSBI metric in the form
\begin{eqnarray}
ds^{2}=-dt^{2}+A^{2}(t)(dx^2+dy^2)+B^{2}(t) dz^2\label{eq1}
\end{eqnarray}
where $A$ and $B$ are the directional scale factors and are considered as functions of cosmic time $t$ only. The metric corresponds to considering xy-plane as the symmetry plane. The eccentricity of such a universe is given by $e=\sqrt{1-\frac{B^2}{A^2}}$. The average scale factor for this metric is  $a=\left(A^{2}B\right)^{\frac{1}{3}}$.

The energy momentum tensor for a pressure less cosmic fluid containing one dimensional strings embedded in an  electromagnetic field is taken as
\begin{eqnarray}
T_{ij}=\rho u_{i}u_{j}-\lambda x_{i}x_{j}+ E_{ij}\label{eq2}
\end{eqnarray}
where $\rho$ is the rest energy density of the system and $\lambda$ is the string tension density. $u^{i}=\delta_{0}^{i}$ are the four velocity vectors. $x^{i}$ is a spacelike vector representing the anisotropic direction of the cosmic strings. $x^{i}$ and $ u^{i}$ satisfy the relations 
\begin{eqnarray}
g_{ij}u^{i}u^{j}=-1,\label{eq3}
\\g_{ij}x^{i}x^{j}=1,\label{eq4}
\\u^{i}x_{i}=0.\label{eq5}
\end{eqnarray}
The one dimensional strings are assumed to spread over the surface of cosmic sheet and aligned along the axis of symmetry. The cosmic strings are  loaded with particles with particle energy density  $\rho_{p}=\rho-\lambda$.
$E_{ij}$ is the part of the energy-momentum tensor corresponding to the electromagnetic field and is given by 
\begin{eqnarray}
E_{ij}=\frac{1}{4\pi}[g^{sp}F_{is}F_{jp}-\frac{1}{4} g_{ij}F_{sp}F^{sp}]\label{eq6}
\end{eqnarray}
where, $F_{sp}$ is the electromagnetic field tensor. We are interested in cosmological models with magnetic field contribution to pressure and therefore, we assume an infinite conductivity of the medium so that only the magnetic components of $F_{sp}$ will exist. We consider a parallel alignment of the magnetic field with respect to the cosmic strings. In otherwords we quantize the axis of magnetic field along the axis of symmetry i.e z-axis. From Maxwell's equations, the only non vanishing component of electromagnetic field tensor comes out to be a constant quantity i.e.
\begin{eqnarray}
F_{12}=-F_{21}=\mathcal{H},\label{eq7}
\end{eqnarray}

where, $\mathcal{H}$ is a constant representing the presence of magnetic field in and around the cosmic strings. If $\mathcal{H}$ is zero, the system is free from any magnetic effect.
\paragraph{} Following closely \cite{Tripathy09a, Mohanty08}, for the plane symmetric metric considered in eq. (1), the components of the electromagnetic field can be expressed as 
\begin{eqnarray}
E_{11}=E_{22}=\frac{\mathcal{H}^{2}}{8\pi A^{2}}=\eta A^{2},\label{eq8}\\
E_{33}=-\frac{\mathcal{H}^{2}B^{2}}{8\pi A^{4}}=-\eta B^{2},\label{eq9}\\
E_{44}=\frac{\mathcal{H}^{2}}{8\pi A^{4}}=\eta.\label{eq10} 
\end{eqnarray}
The field equations in BSCC  
\begin{eqnarray}
G_{ij}=R_{ij}-\frac{1}{2}g_{ij}R=-\frac{8\pi}{\phi} T_{ij}\label{eq11}
\end{eqnarray}
alongwith
\begin{eqnarray}
\square \phi=\frac{8\pi}{3}\zeta T,\label{eq12}
\end{eqnarray}
for the metric in eq.(1), in presence of a cloud of cosmic strings embedded in a magnetic field, can be explicitly expressed as 
\begin{eqnarray}
\left(\frac{\dot{A}}{A}\right)^{2}+2\frac{\dot{A}}{A}\frac{\dot{B}}{B}=\frac{8\pi}{\phi} (\rho+\eta),\label{eq13}
\\\frac{\ddot{A}}{A}+\frac{\ddot{B}}{B}+\frac{\dot{A}}{A}\frac{\dot{B}}{B}=-\frac{8\pi}{\phi}\eta ,\label{eq14}
\\2\frac{\ddot{A}}{A}+\left(\frac{\dot{A}}{A}\right)^{2}=\frac{8\pi}{\phi} (\lambda+\eta),\label{eq15}
\\\ddot{\phi}+\left( 2 \frac{\dot{A}\dot{B}}{AB}\right)\dot{\phi}=\frac{8\pi \zeta}{3} (\rho+\lambda)\label{eq16}
\end{eqnarray}.

In the above field equations, a dot over a directional scale factor represents a time derivative. $\zeta$   is a coupling constant usually  evaluated from experiment and  $\phi$ is the Barber's scalar field.  Here $\phi$  is considered as a function of cosmic time and it encompasses the time varying nature of the Newtonian gravitational constant.  The role of cosmic strings in an anisotropic background has already been investigated in BSCC in Ref. \cite{Mohanty08}. In the present work, in addition to the cloud of cosmic strings, we have incorporated the contribution of magnetic field into the field equations. In the limit $\zeta \rightarrow 0$,   the theory approaches Einstein's general relativity in every respect. For an isotropic flat universe, the directional scale factors are same i.e $A=B=a$ and from eqn \eqref{eq14}and \eqref{eq15}, it is  evident that, in order to get a Friedman like equation, we need to have $\lambda=-2\eta$. In otherwords, for a pressureless cosmic fluid, the contribution of magnetic field and cosmic string comes out as a sort of pseudo anisotropic pressure along the axis of symmetry and the symmetry plane. Magnetic field brings about an anisotropy in the cosmic fluid.

The directional Hubble parameters along the axis of symmetry and symmetry plane are defined as $H_{z}=\frac{\dot{B}}{B}$ and $H_{x}=\frac{\dot{A}}{A}$ so that the mean Hubble parameter becomes $H=\frac{1}{3}\left(2H_{x}+H_{z}\right)$. The scalar expansion and the shear scalar for the metric \eqref{eq1} are respectively expressed as 
\begin{eqnarray}
\theta=2H_{x}+H_{z},\label{eq17}
\\\sigma ^{2}=\frac{1}{3}\left(H_{x}-H_{z}\right)^{2}.\label{eq18}
\end{eqnarray}

Shear scalar is generally considered to be proportional to the scalar expansion which envisages a linear relationship between the directional Hubble parameters i.e.
\begin{eqnarray}
H_{x}=kH_{z}\label{eq19}
\end{eqnarray}
which is equivalent to the anisotropic relationship between the directional scale factors $A$ and $B$ as $A=B^k$. Here $k$ is the anisotropic parameter which should be a positive constant taking care of the anisotropic nature of the model. If $k=1$ the model becomes isotropic otherwise anisotropic. The field equations \eqref{eq13}-\eqref{eq16} can be expressed in terms of the mean Hubble parameter $H$ as 
\begin{eqnarray}
3(k^2+3k+2)\dot{H}+9(k^2+k+1)H^{2}=-\frac{8\pi}{\phi} \left(k+2\right)^{2}\eta, \label{eq20}\\
6(k+2)\dot{H}+27H^{2}=\frac{8\pi}{\phi} \left(k+2\right)^{2} (\lambda+\eta), \label{eq21}\\
9(2k+1)H^{2}=\frac{8\pi}{\phi} \left(k+2\right)^{2}(\rho+\eta). \label{eq22}
\end{eqnarray}

From eqs \eqref{eq20}-\eqref{eq22} we get

\begin{eqnarray}
\frac{8\pi}{\phi}\lambda=\frac{1}{\left(k+2\right)^{2}}\left[3(k^2+5k+6)\dot{H}+9(k^2+k+4)H^{2}\right]\label{eq23}
\end{eqnarray}
and 
\begin{eqnarray}
\frac{8\pi}{\phi}\rho=3\left(\frac{k+1}{k+2}\right)[\dot{H}+3H^{2}] \label{eq24}.
\end{eqnarray}

It is interesting to note from eqn \eqref{eq24} that, the functional  $\chi(H)=\dot{H}+3H^{2}\neq 0$ unlike that in the general relativistic anisotropic model in Ref. \cite{Tripathy13a}. In Ref. \cite{Tripathy13a}, this quantity is zero giving rise to a positive deceleration parameter $q=-1-\frac{\dot{H}}{H^2}$ depicting a decelerating universe. In Ref.\cite{Tripathy13b}, it has been shown  that, a relation of the kind \eqref{eq19} in an LRSBI model in the framework of General Relativity can not predict an accelerating universe as demanded by the host of recent observational data. This is true for whatever matter field taken for the energy momentum tensor excluding electromagnetic contribution. However, the presence of scalar field in generalised Brans-Dicke theory modifies the situation and one can get accelerating model even if the directional Hubble rates are proportional to each other \cite{SKT14}. In the present case i.e, in BSCC, either in the presence or absence of electromagnetic field we will get a non zero value for the functional $\chi(H)$ which shows that there is a possibility of getting an accelerating universe for the ansatz \eqref{eq19} in BSCC. This behaviour of the model may be due to the presence of magnetic field along the direction of the cosmic strings or due to the nature of scalar field. The nature of Barber's scalar field incorporates  the concept of a time varying Newtonian gravitational constant and hence the effect of magnetic field  may be more prominent than the presence of scalar field in getting an accelerating model. 
\paragraph{}For a linear string equation of state $\rho=\gamma\lambda$, the eqs \eqref{eq23}and \eqref{eq24}reduce to 
\begin{equation}
-\frac{\dot{H}}{H^2}=1+q \label{eq25}
\end{equation}
where the deceleration parameter $q$ is expressed as
\begin{equation}
q=2\left[\frac{(k^2-k+3)\gamma-(k^2+3k+2)}{(k^2+5k+6)\gamma-(k^2+3k+2)}\right]\label{eq26}.
\end{equation}
The deceleration parameter comes out to be time independent and depends on the value of the string equation of state and the anisotropic nature  of the model. It is worth to mention here that a positive value of the deceleration parameter favours a decelerating universe whereas its negative value signifies a universe with accelerating phase of expansion. Observations from type Ia Supernovae predict an accelerating universe with deceleration parameter $q=-0.81\pm 0.14$ in the present time \cite{Rapetti07}. Type Ia Supernovae data in combination with BAO and CMB observations constrain the deceleration parameter as $q=-0.53^{+0.17}_{-0.13}$ \cite{Giostri12}. It can be inferred from \eqref{eq26} that, for different choices of the string equation of state $\gamma$, the deceleration parameter assumes either positive or negative values. Baring very low values of the anisotropic parameter $k$, the deceleration parameter becomes negative for all values of $\gamma > 0.36$. In order to get a clear picture about the deceleration parameter, let us now consider two specific choices of the string equation of state i.e the geometric string case with $\gamma=1$ and vacuum string case with $\gamma=-1$.
\paragraph{}For $\gamma=1$, the deceleration parameter becomes
\begin{equation}
q=\frac{1-4k}{2+k}.\label{eq27}
\end{equation}
For an isotropic model, $k=1$ and the deceleration parameter becomes $q=-1$ implying an accelerating universe.  It is clear from \eqref{eq27} that accelerating models can be achieved for all the models with anisotropic parameter greater than 0.25 ie. $k>0.25$. Below this critical value we will get decelerating models.
\begin{figure}[h!]
\begin{center}
\includegraphics[width=1\textwidth]{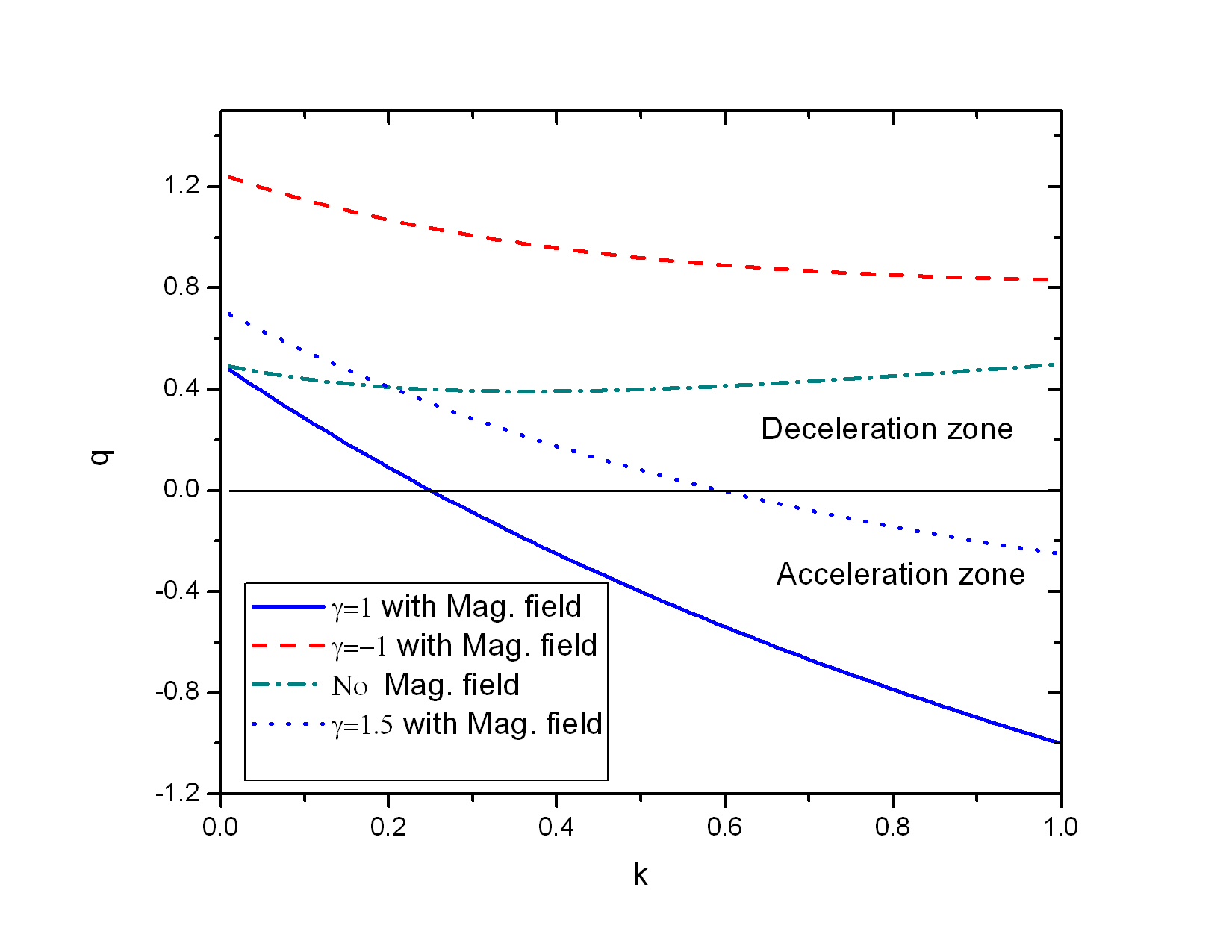}
\caption{Deceleration parameter as a function of anisotropic parameter for two different choices of the string equation of state in presence of magnetic field. Deceleration parameter in the absence of magnetic field is also shown for comparison. $q$ for $\gamma=1.5$ (dotted blue curve)is also shown in the plot.}
\end{center}
\end{figure}
\paragraph{}For the vacuum string case ,i.e. $\gamma=-1$, $\rho+\lambda=0$ i.e. the sum of the rest energy density and string tension density for a cloud of cosmic strings vanishes. Such a situation strips the universe off the string phase and leaves behind an anisotropic cosmic fluid. For this case the deceleration parameter can be expressed as
\begin{equation}
q=\frac{4k^2+k+10}{2(k^2+4k+4)}.\label{eq28}
\end{equation}
Since $k$ is positive, the decelerating parameter comes out to be positive and hence the vacuum string case for the present model can not predict an accelerating universe.  In Fig.1, we have plotted the deceleration parameter as a function of the anisotropic parameter for the two specific choices of the string equation of state. It is clear that, in presence of the magnetic field, the deceleration parameter for $\gamma=1$ decreases with the increase in $k$. For low values of $k$ below $k=0.25$, the deceleration parameter remains in the deceleration zone whereas for higher values of $k$ it goes into the acceleration zone. Basing upon the observational value of the deceleration parameter we can constrain the value of the anisotropic parameter for the geometric string case in the range $0.7<k<0.95$ and $0.69<k<0.72$ corresponding to the values estimated in Refs. \cite{Rapetti07} and \cite{Giostri12} respectively. For $\gamma=-1$, the deceleration parameter also decreases with the increase in anisotropic parameter but remains completely in the deceleration zone. In other words, for the present anisotropic model, it is unreasonable to think of a vacuum string case which does not support the host of observational data predicting an accelerating universe.

For any general value of the string equation of state parameter $\gamma$ in the positive regime, particularly for $\gamma >0.36$, the deceleration parameter decreases from positive value for small values of $k$ to negative values for higher $k$ after crossing the deceleration to acceleration transit boundary. In order to get an idea about the general scenario for $\gamma$, we have shown the plot for $\gamma =1.5$ (dotted blue curve in Fig.1). This curve moves down from the deceleration zone to acceleration zone with an increase in $k$ but the transit boundary crossing occurs at a higher value of $k$ than that for $\gamma=1$ (blue solid curve). With the increase in the value of $\gamma$, the crossing point shifts more and more towards a higher value of $k$. For much higher values of $\gamma$, the curves for $q$ will remain above the transit boundary.

\paragraph{}Integration of Eq.\eqref{eq25} yields
\begin{equation}
H=\frac{H_0}{(1+q)H_{0}(t-t_0)+1},\label{eq29}
\end{equation}
where $H_0$ is the Hubble parameter in the present time $t_0$. Consequent upon Eq. \eqref{eq29}, we can calculate the scale factor as
\begin{equation}
a=a_{0}\left[(1+q)H_{0}(t-t_0)+1\right]^{\frac{1}{1+q}},\label{eq30}
\end{equation}
where $a_0$ is the scale factor in the present epoch. The redshift $z$ can also be calculated using the fact $1+z=\frac{a_0}{a}$ as 
\begin{equation}
z=-1+\left[(1+q)H_{0}(t-t_0)+1\right]^{-\frac{1}{1+q}}.\label{eq31}
\end{equation}
From Eqs. \eqref{eq10} and \eqref{eq30}, we can have the expression of  the normalized energy density contribution from magnetic field as 
\begin{equation}
\eta=\eta_0\left[(1+q)H_{0}(t-t_0)+1\right]^{-\frac{12}{(1+q)(k+2)}}, \label{eq32}
\end{equation}
where $\eta_0=\frac{\mathcal{H}^2}{8\pi A_{0}^4}$ and $A_0$ correspond to the values at  the present time. The magnetic energy density decreases with the growth of cosmic time and at late time of evolution its effect becomes negligible.

\subsection{Physical and geometrical properties of the model}
The rest energy density ($\rho$), the string tension density($\lambda$), the particle density ($\rho_p$) for the model are given by

\begin{equation}
\rho=\rho_{0}\left[(1+q)H_{0}(t-t_0)+1\right]^{-\frac{12}{(1+q)(k+2)}},\label{eq33} 
\end{equation}
\begin{equation}
\lambda=\lambda_0\left[(1+q)H_{0}(t-t_0)+1\right]^{-\frac{12}{(1+q)(k+2)}},\label{eq34}
\end{equation}

\begin{equation}
\rho_p=\left(\frac{\gamma-1}{\gamma}\right)\rho_{0}\left[(1+q)H_{0}(t-t_0)+1\right]^{-\frac{12}{(1+q)(k+2)}},\label{eq35}
\end{equation}
where 
\begin{equation}
\rho_0=\frac{(k+1)(2-q)(k+2)\eta_0}{3(k^2+K+1)-(1+q)(k^2+3k+2)},\label{eq36}
\end{equation}
and
\begin{equation}
\lambda_0=\frac{(k+1)(2-q)(k+2)\eta_0}{\gamma\left[(k^2+K+1)-(1+q)(k^2+3k+2)\right]},\label{eq37}
\end{equation}
are the rest energy density and string tension density in the present epoch $t_0$.
\paragraph{}The Barber Scalar field can be expressed as
\begin{equation}
\phi=\phi_0\left[(1+q)H_{0}(t-t_0)+1\right]^{2-\frac{12}{(1+q)(k+2)}}.\label{eq38}
\end{equation}
In Eq. \eqref{eq38}, $\phi_0=\frac{(k+2)^2}{9(k^2+k+1)-3(1+q)(k^2+3k+2)}\frac{8\pi\eta_0}{H_{0}^2}$ is the value of the scalar field in the present time.
\paragraph{}The rest energy density and the string tension density decreases with the expansion of the universe. In the beginning of the universe, they have large magnitudes and they roll down to small values at late time of evolution. The increment or decrement of the scalar field with time depends on the choice of the anisotropic parameter and the string equation of state and hence the deceleration parameter. For $(1+q)(k+2)<6$, the scalar field decreases with time whereas for $(1+q)(k+2)>6$, it increases with time. At the critical value $(1+q)(k+2)=6$, the scalar field becomes independent of time. Considering the present observational estimates of deceleration parameter \cite{Rapetti07, Giostri12}and the plausible constraints on the range of anisotropic parameter $k$, $(1+q)(k+2)$ will always be less than 6 and hence the scalar field decreases with the growth of cosmic time.
\paragraph{}The geometrical features of the model are expressed through the shear scalar $\sigma^2$ and the scalar expansion $\Theta$. For the present model these quantities are given by
\begin{eqnarray}
\sigma^2=3\left(\frac{k-1}{k+2}\right)^{2}\left[\frac{H_0}{(1+q)H_{0}(t-t_0)+1}\right]^2,\\
\Theta=\frac{3H_0}{(1+q)H_{0}(t-t_0)+1}.
\end{eqnarray}
The shear scalar and the scalar expansion decrease from large value at the beginning of cosmic time to small value at late times.
\subsection{Cosmological model in the absence of magnetic field}
In the absence of magnetic field, the field equations in Self Creation Cosmology assume the form
\begin{eqnarray}
3(k^2+3k+2)\dot{H}+9(k^2+k+1)H^{2}=0, \label{eq39}\\
6(k+2)\dot{H}+27H^{2}=\frac{8\pi}{\phi} \left(k+2\right)^{2} \lambda, \label{eq40}\\
9(2k+1)H^{2}=\frac{8\pi}{\phi} \left(k+2\right)^{2}\rho. \label{eq41}
\end{eqnarray}
From Eq.\eqref{eq39}we can have
\begin{equation}
-\frac{\dot{H}}{H^2}=1+\frac{2k^2+1}{k^2+3k+2},\label{eq42}
\end{equation}
so that the deceleration parameter $q=-1-\frac{\dot{H}}{H^2}$ is given by
\begin{equation}
q=\frac{2k^2+1}{k^2+3k+2}.\label{eq43}
\end{equation}
It is interesting to note here that, since the anisotropic parameter $k$ is always positive, the deceleration parameter is also a positive constant quantity independent of cosmic time. It only depends on the  anisotropic nature of the model. The deceleration parameter in the absence of magnetic field first decreases then increases with the increase in the anisotropic parameter (see Fig.1). However, its variation with respect to $k$ is very less. It remains totally in the deceleration zone.  It is amply clear from the calculation that in the absence of magnetic field, in the present model, it is not possible to get an accelerated phase of expansion of the universe. A similar conclusion has also been derived in Ref. \cite{Tripathy13b} where it has been shown that in Einstein's general relativity, a relation of the type \eqref{eq19} in the absence of magnetic field can not predict an accelerating universe which necessitates either the consideration of magnetic field in the field equation or a more evolving relationship among the directional Hubble parameters. However, in the present case of Self Creation Cosmology, the role of magnetic field in getting an accelerating universe is clearly established for a linear relationship among the directional Hubble parameters.
\paragraph{}We can get the string equation state from Eqs \eqref{eq40}and \eqref{eq41} as
\begin{equation}
\gamma=\frac{\rho}{\lambda}=1+\frac{2k(2k+1)}{1+k-2k^2}.\label{eq44}
\end{equation}
\begin{figure}[h!]
\begin{center}
\includegraphics[width=1\textwidth]{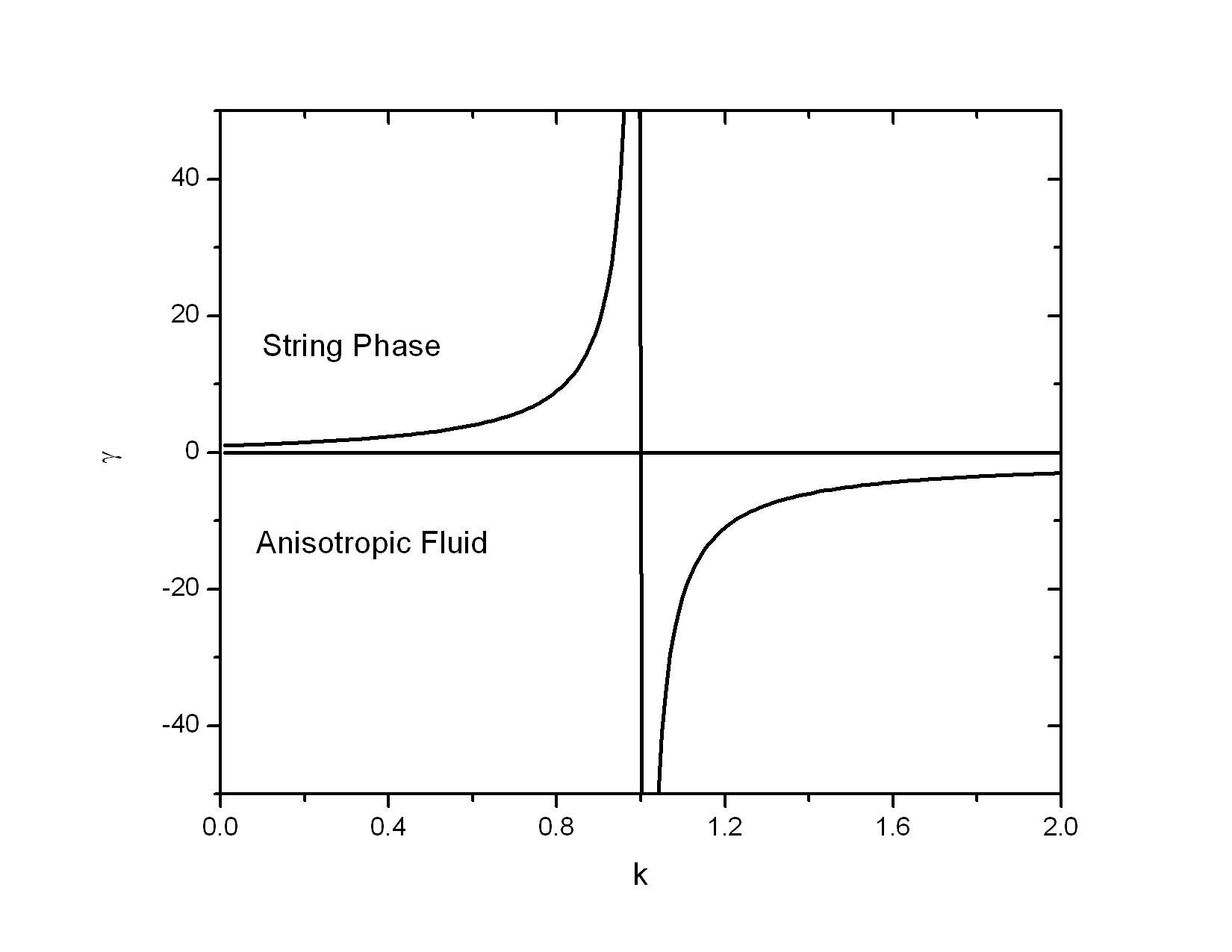}
\caption{String equation of state  as a function of anisotropic parameter in the absence of magnetic field(See eq. (46)).}
\end{center}
\end{figure}
\paragraph{} In Fig.2, the string equation of state in the absence of magnetic field is plotted as a function of anisotropic parameter. The string equation of state is independent of time and depends only on the value of the anisotropic parameter $k$. However, this expression \eqref{eq44}is not valid for an isotropic model with $k=1$. The cosmic string in the absence of magnetic field favours a Takabayasi string equation of state for all values of $k$ satisfying the relation $1+k>2k^2$. It is worth to mention here that, a Takabayasi string equation of state is represented by $\rho=(1+W)\lambda$ or $\gamma = 1+W$, where, $W>0$ is a constant \cite{Taka74,Taka74a,Ruggeri78}. For $W=0$, the Takabayasi string reduces to the usual geometric string case with $\gamma=1$, usually referred to as Nambu string. In other cases, for all positive values of $k$, the string equation of state favours a relationship $\rho <\lambda$. For $k<1$, the string equation of state is positive whereas for $k>1$,  $\gamma$ is negative. In other words, we can get a string phase in the universe for anisotropic parameter in the range $k<1$. 

In the model with cosmic strings embedded in a parallel magnetic field (Sec-2), we have restricted ourselves to a linear string equation of state to obtain the analytical solutions to the field equations. But if we try to figure out the exact relationship between the rest energy density and the string tension density, we may get a different picture than the linear one.

\paragraph{} In presence of magnetic field, the string equation of state can be expressed as
\begin{equation}
\gamma=\frac{\rho}{\lambda}=\frac{(k+1)(k+2)\left(\dot{H}+3H^2\right)}{\left[(k^2+5k+6)\dot{H}+3(k^2+k+4)H^{2}\right]}.\label{eq45}
\end{equation}
Since the functional $\chi(H)=\dot{H}+3H^{2}\neq 0$ for the present model with magnetic field having a positive rest energy density, eq.$\eqref{eq45}$ clearly indicates an evolving relationship between the rest energy density and the string tension density provided that the time dependence of the numerator in $\eqref{eq45}$ is not exactly cancelled by the time dependence of the denominator. In other words, in presence of magnetic field, the relationship among the rest energy density and string tension density evolves all through the expansion history of the universe. However, for a linear time independent string equation of state, the functional $\chi(H)$ reduces to the one described by eq.\eqref{eq25}.

\section{Models with Dark Energy components}
The presence of exotic matter and/or energy in the universe is believed to provide an acceleration. The driving force is the anti gravity affect of the dark energy that generates a strong negative pressure.  We consider the presence of components of dark energy in the anisotropic model in the form of a time varying cosmological constant term $\Lambda$ in the field equations. Recent cosmological observations suggest a small but positive cosmological constant with magnitude $\Lambda(\frac{G\hbar}{c^3})=10^{-123}$ \cite{Reiss98, Perl98}.
\paragraph{} The field equations in the frame work of Self Creation Cosmology in presence of magnetic field and dark energy components become
\begin{eqnarray}
3(k^2+3k+2)\dot{H}+9(k^2+k+1)H^{2}=-\frac{8\pi}{\phi} \left(k+2\right)^{2}\eta+\Lambda, \label{eq46}\\
6(k+2)\dot{H}+27H^{2}=\frac{8\pi}{\phi} \left(k+2\right)^{2} (\lambda+\eta)+\Lambda, \label{eq47}\\
9(2k+1)H^{2}=\frac{8\pi}{\phi} \left(k+2\right)^{2}(\rho+\eta)+\Lambda. \label{eq48}
\end{eqnarray}

For a general string equation of state satisfying $\rho=\gamma \lambda$, the field equations $\eqref{eq46}-\eqref{eq48}$ reduce to 
\begin{equation}
6(k+2)\dot{H}-18(k-1)H^2=\frac{8\pi}{\phi} \left(k+2\right)^{2} (1-\gamma)\lambda \label{eq54}.
\end{equation}

It can be noticed from the discussion in Sec-2.1 that, the ratio of the string tension density and the scalar field behaves like the square of the Hubble parameter i.e. $\frac{\lambda}{\phi} \sim H^2$. Assuming a similar relationship in the present model, we can obtain a time independent deceleration parameter  as

\begin{equation}
q=\frac{1-4k}{2+k}+\frac{\xi (k,\gamma)}{6(k+2)},\label{eq55}
\end{equation}
where, $\xi (k,\gamma)=8n\pi (k+2)(1-\gamma)$ and $n=\frac{\lambda}{\phi H^2}$. The functional $\xi$ vanishes for geometric string case with $\gamma=1$. For positive values of $n$, $\xi$ is negative for $\gamma>1$ and is positive for $\gamma <1$. The deceleration parameter is time independent and depends on the choice of the anisotropic parameter $k$ and the string equation of state parameter $\gamma$. Since the geometric string case provides  interesting results in cosmological model with magnetic field ( Sec.2) without dark energy component, in this section, we will discuss only this particular case. It is straightforward to calculate the deceleration parameter for  geometric cosmic string ($\gamma=1$) as
\begin{equation}
q=\frac{1-4k}{2+k}.\label{eq49}
\end{equation}
This expression is the same as that of $\eqref{eq27}$ and the deceleration parameter behaves in the manner as depicted in Fig.1 for $\gamma =1$ (blue solid curve). The deceleration parameter depends only on the anisotropic parameter and is time independent. It assumes negative values for the range $k>0.25$. In otherwords, the presence of the dark energy component along with the magnetic field favours an accelerating universe in the range $k>0.25$. For $\gamma <1$, the deceleration parameter will be more than that of $\eqref{eq49}$ and it will be less for  $\gamma >1$.

\paragraph{} The cosmological constant is now expressed by
\begin{equation}
\Lambda=\Lambda_0\left[(1+q)H_{0}(t-t_0)+1\right]^{-2},\label{eq50}
\end{equation}
where $\Lambda_0=9(2k^2+k)H_{0}^2
+(k^2+2)^{2}8\pi \frac{\eta_{0}}{\phi_{0}}$ is the value of the cosmological constant in the present epoch. In conformity with the experimental evidences, the cosmological constant decreases with time from large value at an initial epoch to small positive value at late time of evolution.
\paragraph{} In the absence of magnetic field, the field equations \eqref{eq46}-\eqref{eq48} predict the same deceleration parameter for geometric cosmic strings spreading the surface of the world sheet as that of \eqref{eq49}.\\ The cosmological constant in the absence of magnetic field becomes 
\begin{equation}
\Lambda=\Lambda_0\left[(1+q)H_{0}(t-t_0)+1\right]^{-2}.\label{eq51}
\end{equation} 
The cosmological constant in the absence of magnetic field takes the same form of (54) but now with a different value at the present epoch, $\Lambda_0=9(2k^2+k)H_{0}^2$. It is interesting to note that even in the absence of magnetic field it is possible to get an accelerating model if we incorporate dark energy component in the field equations. Even though, the exact order of magnitude of contribution from magnetic field in the present model can not be estimated, it is certain from the above discussion that, the role of magnetic field in getting an accelerating model is overshadowed by the dark energy component. However, one can not rule out the role played by magnetic field in the present model even in presence of dark energy components. The presence of magnetic field brings about a change in the magnitude of the cosmological constant in the present epoch. In fact, the cosmological constant in the present epoch is somewhat lowered in absence of magnetic field.

\begin{figure}[h!]
\begin{center}
\includegraphics[width=1\textwidth]{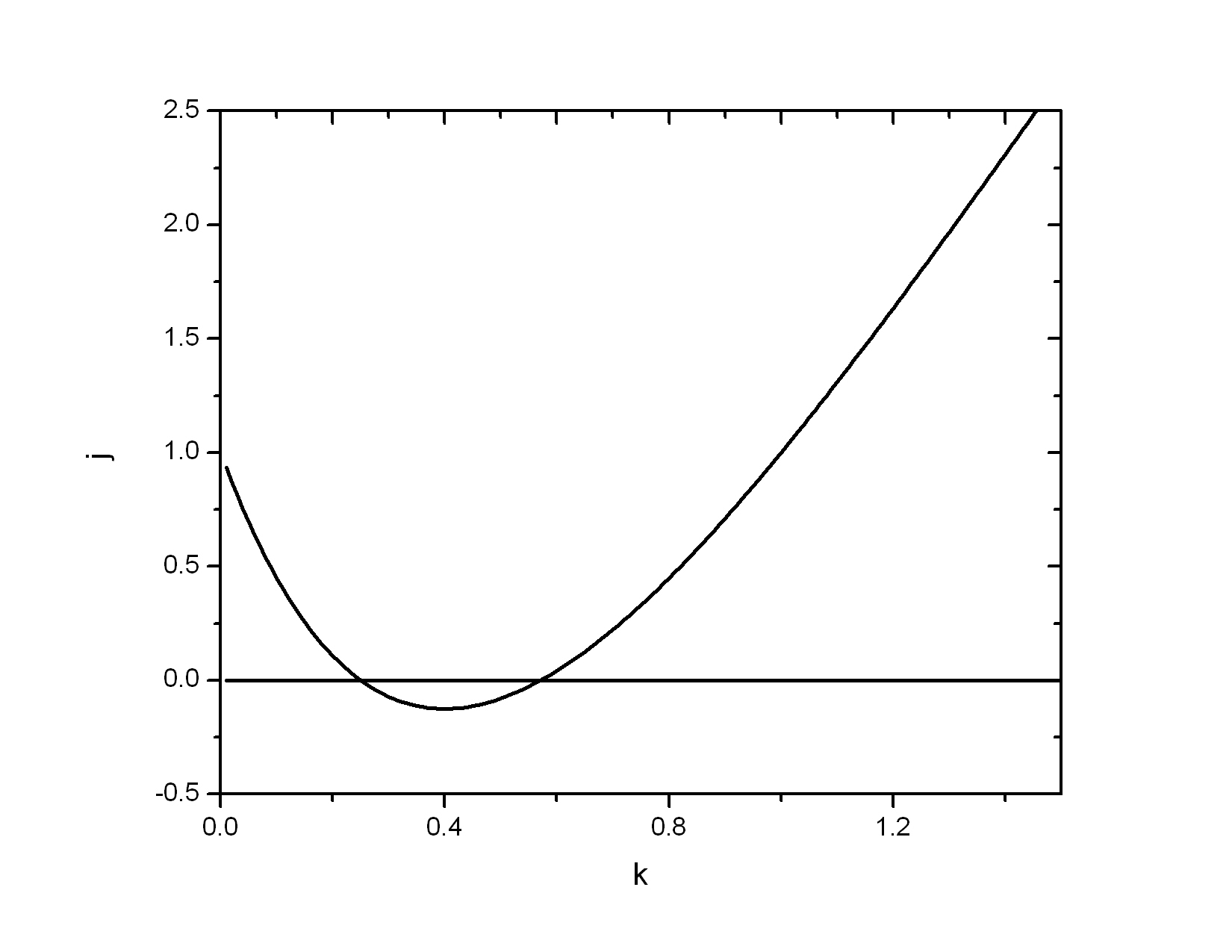}
\caption{Jerk parameter  as a function of anisotropic parameter.}
\end{center}
\end{figure}

\paragraph{} The jerk parameter for this dark energy model is expressed as
\begin{equation}
j=\frac{\dddot a}{a H^3}=\frac{28k^2-23k+4}{(k+2)^2}.\label{eq52}
\end{equation}
The jerk parameter comes out to be a constant quantity and depends on the choice of the anisotropic parameter. In Fig.3, jerk parameter is plotted as a function of anisotropic parameter .  It is positive for all values of the anisotropic parameter $k$ except for the range $1/4<k<4/7$. In this range $j$ becomes negative. The jerk parameter decreases with the increase in the anisotropic parameter up to $k=0.41$ and then increases. For an isotropic case with $k=1$, the jerk parameter becomes $j=1$ in conformity with the prediction from $\Lambda$CDM model. In the present model, since the anisotropic parameter is constrained to be in the range $k<1$, $j$  can take values less than one encompassing values from both the positive and negative domain. It is worth to mention here that the exact determination of the jerk parameter requires the observation of Supernovae of redshift greater than one which is presently a difficult task and therefore, current observational data invoking type Ia Supernovae are not able to pin down the value or the sign of the jerk parameter. If observational data from other sources such as CMB and BAO can provide a good estimate of the cosmological parameters i.e deceleration parameter and jerk parameter, then accordingly it is possible to constrain the anisotropic parameter $k$ to a more tighter range.

\section{Conclusion}
\paragraph{}The universe is not only expanding but the expansion is accelerating. The reason behind the acceleration is not yet known to a satisfactory extent. In the present work, we have investigated the role of magnetic field in presence of cloud of cosmic strings pervading the world sheet to get accelerating models. For this purpose, we considered plane symmetric cosmological models in the frame work of Barber's Self Creation Cosmology. In order to get determinate cosmological models we have assumed a linear relationship between the directional Hubble parameters which envisages an anisotropic relationship among the directional scale factors. In an earlier work \cite{Tripathy13a}, it has been shown that a linear relationship among the directional Hubble parameters will not be able to predict the observational facts concerning the accelerating phase of the universe. However, in the present work,  we get accelerating models even if we consider such anisotropic relationship. This type of behaviour may be due to the presence of magnetic field or due to the nature of scalar field. In the absence of magnetic field, the deceleration parameter comes out to be positive for all possible values of anisotropic parameter whereas in presence of magnetic field the deceleration parameter can be negative for a range of anisotropic parameter implying an accelerating universe. Even if it is not possible to extract an estimate of the order of contribution from magnetic field in our model, it can be assessed that, magnetic field plays some interesting role in the cosmic dynamics. However, incorporation of dark energy components into the field equations in the form of a time varying cosmological constant hides the role of magnetic field. Basing upon the experimental values of deceleration parameter, we have tried to constrain the anisotropic parameter $k$ for geometric string case within a very narrow range. In future, if the deceleration parameter and the jerk parameter can be determined from observations with acceptable accuracy, then we hope, from our simple model, the spatial anisotropy in the form of anisotropic parameter can be well pinned down.

\end{document}